\documentclass[twocolumn,showpacs,preprintnumbers,superscriptaddress,amsmath,amssymb]{revtex4}

\usepackage{graphicx}
\usepackage{dcolumn}
\usepackage{bm}

\begin{document}

\preprint{APS/123-QED}

\title{Theoretical study of resonant x-ray emission spectroscopy of Mn films on Ag}

\author{M. Taguchi}

\affiliation{Soft X-ray Spectroscopy Lab, RIKEN/SPring-8, Sayo, Sayo, Hyogo 679-5148, Japan}

\author{P. Kr\"uger}

\affiliation{Laboratoire de Recherches sur la R\'eactivit\'e des Solides (LRRS) UMR 5613 Universit\'e de Bourgogne - CNRS
Boite Postale 47870, 21078 Dijon, France}

\author{J. C. Parlebas}

\affiliation{Institut de Physique et de Chimie des Mat\'eriaux de Strasbourg, UMR 7504, CNRS, 23, rue du Loess, 67034 Strasbourg Cedex 02, France}

\author{A. Kotani}

\affiliation{Soft X-ray Spectroscopy Lab, RIKEN/SPring-8, Sayo, Sayo, Hyogo 679-5148, Japan}

\affiliation{Photon Factory, IMSS, High Energy Accelerator Research Organization, Tsukuba, Ibaraki 305-0801, Japan}

\date{\today} 

\begin{abstract}
We report a theoretical study on resonant x-ray emission spectra (RXES) in the whole energy region of the Mn $L_{2,3}$ white lines for three prototypical Mn/Ag(001) systems: (i) a Mn impurity in Ag, (ii) an adsorbed Mn monolayer on Ag, and (iii) a thick Mn film. The calculated RXES spectra depend strongly on the excitation energy.  At $L_3$ excitation, the spectra of all three systems are dominated by the elastic peak.  For excitation energies around $L_2$, and
between $L_3$ and $L_2$, however, most of the spectral weight comes from
inelastic x-ray scattering. The line shape of these inelastic ``satellite''
structures changes considerably between the three considered Mn/Ag systems, a fact that may be attributed to changes in the bonding nature of the Mn-$d$ orbitals.  The system-dependence of the RXES spectrum is thus found to be much stronger than that of the corresponding absorption spectrum. Our results suggest that RXES in the Mn $L_{2,3}$ region may be used as a sensitive probe of the local environment of Mn atoms. 
\end{abstract}


\maketitle

\section{Introduction}

With the advent of 3rd generation sources of synchrotron radiation, resonant x-ray emission spectroscopy (RXES) has recently been used to investigate the nature of localization of the $3d$ and/or $4f$ electrons for strongly correlated electron systems\cite{kot01,but00}.  
In particular, RXES has often been applied to transition metal (TM) compounds
in order to study intra-atomic ($dd$) excitations as well as charge transfer
(CT) excitations\cite{kot01,but00, mag02,ghi04}. An accurate mapping of these two types of excitations
is very useful for a better understanding of these systems.
In $2p\to3d\to2p$ RXES spectra of typical TM compounds such as MnO,
the features of the CT excitation are clearly separated from those of
the $dd$ excitation, because in these systems the CT energy $\Delta$ is large\cite{but96}.
This separation obviously greatly simplifies the analysis of the spectra.
It is, however, important to explore the behavior of these
two types of excitations also in the less favorable case
where a clear separation of $dd$ and CT features in the RXES spectra is
impossible. It is the first aim of this paper to present such a study.

The second aim is to make predictions for RXES spectra of Mn films on
Ag(001), which is an interesting and well characterized example of
supported ultra thin transition metal films.
The fundamental and technological interest of these systems, in particular
their magnetic properties, is well known \cite{dru88,rau88,bin92,kru96,rad97,sch97,kru99}.

Recently, we have presented calculations of the $2p$ x-ray photoemission
(XPS), $2p$ x-ray absorption (XAS), and (a few)
resonant x-ray emission spectra (RXES) of several Mn/Ag thick film
structures\cite{kru03,tag04}.  
By combining band structure, atomic multiplet and impurity Anderson model calculations, we constructed a realistic impurity model that includes full intra-atomic multiplet interaction and inter-atomic coupling to the Mn-$3d$ and Ag-$4d$ bands. The calculated Mn $2p$ XPS spectra reproduced well the experimental ones in the whole range of structures from Mn impurities in Ag to bulk bct Mn. These calculations indicate that the satellite structure observed in the ultra-thin films is due to a final state charge transfer process, mainly from the Mn-$d$ majority spin state on the neighboring atoms to the empty minority spin state of the emitter atom. 
As an interesting "by-product" of these studies we found that the CT energy
$\Delta$ is much smaller in Mn/Ag than in MnO.
This fact suggests that in Mn/Ag, CT and $dd$ excitations coexist in the same
energy range. Mn/Ag is therefore an interesting case for our purpose,
all the more that in these thick film systems, two parameters relevant
for the CT excitation (CT energy and hybridization) can, to some extent,
be controlled by the Mn coverage.

In order to vary these parameters, we have selected the following
three Mn/Ag(001) systems for the present study:
(i) a Mn impurity atom featuring weak hybridization to the Ag $d$-band
with moderate CT energy,
(ii) an adsorbed Mn monolayer featuring weak and moderate hybridization to Ag and Mn $d$-bands, respectively, with moderate and small CT energy, and
(iii) a thick Mn film characterized by strong hybridization to the
Mn $d$-band with small CT energy.

Our previous paper in Ref.\cite{tag04} was devoted to a
comparative study of XPS, XAS, RXES of Mn/Ag in order to construct a
single model that is valid for all these spectroscopies.
In the RXES part, only a few examples of spectra were shown. 
The present paper, in contrast, is devoted to a systematic study
of the Mn $2p\to3d\to2p$ RXES in a wide range of the incident photon energy
for the three above mentioned Mn/Ag systems.
We analyze the competition between CT and $dd$ excitations in RXES
and compare the different systems among them and with MnO.
We also study the dependence of the RXES line shape on the
hybridization strength.
One main result is that the Mn $2p\to3d\to2p$ RXES line shape
varies appreciably as a function of hybridization and CT energy
and the RXES spectra differ, accordingly, strongly between the
three different Mn/Ag systems.
This suggests that $2p\to3d\to2p$ RXES may be used as a sensitive probe
for the local environment of transition metal atoms.

 The paper is organized as follows. In section II, we discuss the theoretical model. The results and discussion are given in section III and VI, respectively.

\section{Method of calculation}

\begin{figure}
\begin{center}
\includegraphics[scale=.53]{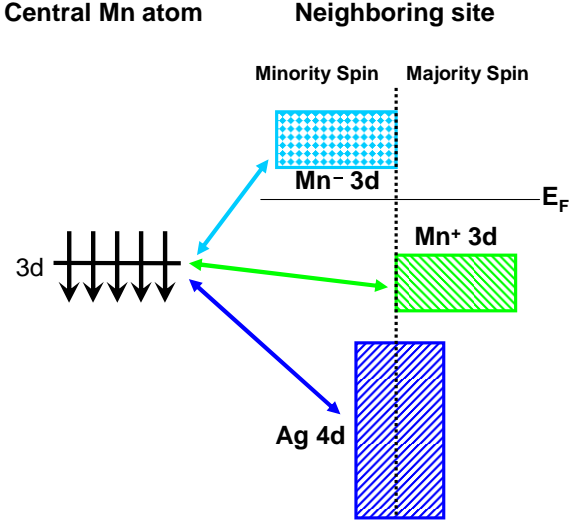}

\caption{\label{fig1} (Color online) Schematic diagram of the impurity model for Mn/Ag system. The Fermi level (E$_{\rm F}$) separates the occupied density of states from the unoccupied density of states.}
\end{center}
\end{figure}

Here we give a brief description of the model. Further details may be found in Refs.~\cite{kru03} and \cite{tag04}. For the system one Mn monolayer on Ag, a schematic diagrum of our impurity model is given in Fig.~1 which is based on the results of $ab$ $initio$ calculation\cite{kru03}.  
We take into account the Mn $3d$ orbital of a central Mn atom (core hole site, which has a $3d^5$ configuration) together with appropriate linear combinations of Mn $3d$ orbitals (while the Mn $3d$ band is filled for locally majority spin, which is denoted as Mn$^+$, and empty for locally minority spin, which is denoted as Mn$^-$) on neighboring sites and Ag $4d$ filled orbitals of the Ag substrate. 

\begin{table}
\begin{tabular}{cccccccccc}
\multicolumn{10}{c}{system independent impurity model parameters }\\
\hline
& $\Delta$ & $U_{dd}$ & $U_{dc}$ & $\varepsilon_{\rm Mn^+}$ & $\varepsilon_{\rm Mn^-}$ & $\varepsilon_{\rm Ag}$ & $N$ & $10Dq$ & \\
& -3.012 & 3.0 & 4.123 & -2.5 & 0.5 & -5.0 & 4 &0.5 & \\
\hline
\multicolumn{10}{c}{system dependent impurity model parameters }\\
\hline
\multicolumn{4}{l}{system} & $W(\rm Ag)$ & $t_{\rm Ag}(e_g)$ & $W(\rm Mn)$ & $t_{\rm Mn}(e_g)$ & \\
\multicolumn{4}{l}{impurity in Ag} & 3.0 & 0.866 & 0.0 & 0.0 & \\
\multicolumn{4}{l}{ML} & 3.0 & 0.5 & 1.0 & 0.8 & \\
\multicolumn{4}{l}{Mn bct} & 0.0 & 0.0 & 3.0 & 2.0 & \\
\hline
\end{tabular}
\caption{Energy parameter values (in eV) used in the calculations.  }
\end{table}

The thick Mn film is modeled as bulk bct Mn.
A single impurity Mn atom in Ag is hybridized to the Ag $4d$-band only.

The ground state of Mn monolayer on Ag system is represented as linear combinations of basis states from the manifold of the following eight configurations: $3d^5$, $3d^{5+n}(\underline{\rm Ag})^n$, $3d^{5+n}(\underline{\rm Mn^+ })^n$, $3d^4(\rm Mn^- )$, $3d^5(\rm Mn^-)(\underline{\rm Mn^+})$, $3d^7(\underline{\rm Mn^+})(\underline{\rm Ag})$, (n=1,2). $\underline{\rm Ag}$ and $\underline{\rm Mn^+}$ denote holes in the Ag band and in the Mn band with (locally) majority spin, respectively. Mn$^-$ represents an electron in the Mn band with (locally) minority spin. 
For bct thick films, we have used the five configuration ($3d^5$, $3d^{5+n}(\underline{\rm Mn^+ })^n$, $3d^4(\rm Mn^- )$, $3d^5(\rm Mn^-)(\underline{\rm Mn^+})$, n=1,2) and for Mn single atom on Ag system, only three configurations ($3d^5$, $3d^{5+n}(\underline{\rm Ag})^n$, n=1,2) were used. We assume a cubic point symmetry for all systems for simplicity\cite{tag04}. Note that, in Mn monolayer on Ag and Mn bct systems the local symmetry around the Mn atom is approximately O$_h$, and that in Mn impurity in Ag is exactly O$_h$.

The Hamiltonian is given by 

\begin{eqnarray}
H  &=&  \sum_{\Gamma,\sigma} \varepsilon_{3d}(\Gamma)d^{\dagger}_{\Gamma\sigma}d_{\Gamma\sigma}
 \hspace{-1pt}+ \hspace{-1pt}\sum_{m,\sigma}\varepsilon_{2p}p^{\dagger}_{m \sigma}p_{m \sigma} \nonumber \\
    &+& U_{dd}\sum_{(\Gamma,\sigma)\neq(\Gamma',\sigma')}d^{\dagger}_{\Gamma\sigma}
d_{\Gamma\sigma}d^{\dagger}_{\Gamma'\sigma'}d_{\Gamma'\sigma'}\hspace{2.0cm} \nonumber \\
  &-& U_{dc}\sum_{\Gamma,m,\sigma,\sigma'}d^{\dagger}_{\Gamma\sigma}d_{\Gamma\sigma}
(1 - p^{\dagger}_{m\sigma'}p_{m\sigma'}) + H_{\rm{multiplet}} \nonumber \\ 
  &+& \sum_{X,k,\sigma}\varepsilon_{X}(k)a^{\dagger}_{Xk\sigma}a_{Xk\sigma} \nonumber \\
  &+& \sum_{X,k,\Gamma,\sigma}\frac{t_X(\Gamma)}{\sqrt{N}}(d^{\dagger}_{\Gamma\sigma}a_{Xk\sigma}
 + a^{\dagger}_{Xk\sigma}d_{\Gamma\sigma}) \ \ .
\end{eqnarray}

The first five terms of the total Hamiltonian~$H$
are the atomic part describing the central Mn atom.
The sixth term represents the $X$ band ($X=\rm Mn^+$, $\rm Mn^-$ and Ag) 
and the last term describes the hybridization between the atomic Mn $3d$ 
states and the $X$ band. 
The $\varepsilon_{3d}(\Gamma)$, $\varepsilon_{2p}$, and $\varepsilon_{X}(k)$ 
represent the energies of Mn 3$d$, Mn 2$p$, and $X$ band states, respectively, 
with the irreducible representation $\Gamma$ ( = $e_g$, $t_{2g}$) of the 
O$_h$ symmetry. 
The indices $m$ and $\sigma$ are the orbital and spin states. 
$V_X(\Gamma)$, $U_{dd}$, and $U_{dc}$ are the hybridization 
between the central Mn 3$d$ and $X$ band states, 
the Coulomb interaction between Mn 3$d$ states, 
that between Mn 3$d$ and 2$p$ core-hole states, respectively. 
The term $H_{\rm{multiplet}}$ describes the intra-atomic multiplet coupling 
between Mn 3$d$ states and that between Mn 3$d$ and 2$p$ states. 
The spin-orbit interactions for Mn 3$d$ and 2$p$ states are also included. 
The Slater integrals and the spin-orbit coupling constant are calculated by 
Cowan's Hartree-Fock program\cite{cow81} and then the Slater integrals are 
rescaled by 80$\%$, as usual\cite{gro90}. 
We assume rectangular bands $X$ of width W(X), and we approximate them
by $N$ discrete levels 
$\varepsilon_X(k)=\varepsilon_{X} + \frac{W(X)}{N}(k-\frac{N+1}{2})$,
($k=1,...,N$). 
For the hybridization, we use the empirical relation: $V_X(e_g)=-2V_X(t_{2g})$.
 The charge transfer energy $\Delta$ is defined as 
$\Delta \equiv \varepsilon_{3d} +5 U_{dd}$.

\begin{figure}
\begin{center}
\includegraphics[scale=.53]{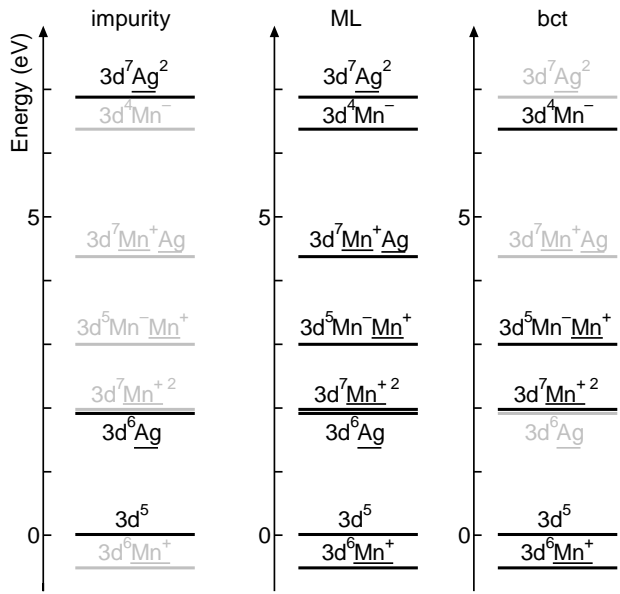}

\caption{\label{fig2} Schematic energy diagram of the ground state configurations for the different systems (see text). }
\end{center}
\end{figure}

The main contribution to Mn $2p\to3d\to2p$ RXES corresponds to the process $2p^63d^n\to 2p^53d^{n+1}\to 2p^63d^n$ where $n$ is the $3d$ occupation number in the ground state. 
Mn $2p\to3d\to2p$ RXES is calculated on the basis of the formula of a coherent second order optical process as 

\begin{eqnarray}
F(\Omega,\omega) &=&
 \sum_{f} \biggl|\sum_{m} \frac{ \langle f \mid T^{\dagger} \mid m \rangle \langle m 
\mid T \mid g \rangle}{E_{g}+ \Omega-E_{m}-i\Gamma_{L}}\biggr|^{2} \nonumber \\
&\times&\delta(E_{g}+ \Omega - E_{f}- \omega)   \ \ ,
\end{eqnarray}

\noindent
where $\mid g \rangle$, $\mid m \rangle$ and $\mid f \rangle$ are the ground, 
intermediate and final states of the Hamiltonian $H$ with energies $E_g, E_m$ and $E_f$, respectively. 
The incident and emitted photon energies are represented by $\Omega$ and $\omega$, respectively. 
The core-hole lifetime broadening is denoted by $\Gamma_L$ for the $2p$ core-hole in the intermediate states. The operators $T$ represents the optical dipole transition. The polarization of the incident photon is neglected, for simplicity.

Unless stated explicitly, the parameter values are the same as in Ref.~\cite{tag04} and are also summarized in Table~I.

\begin{figure}
\begin{center}
\includegraphics[scale=.50]{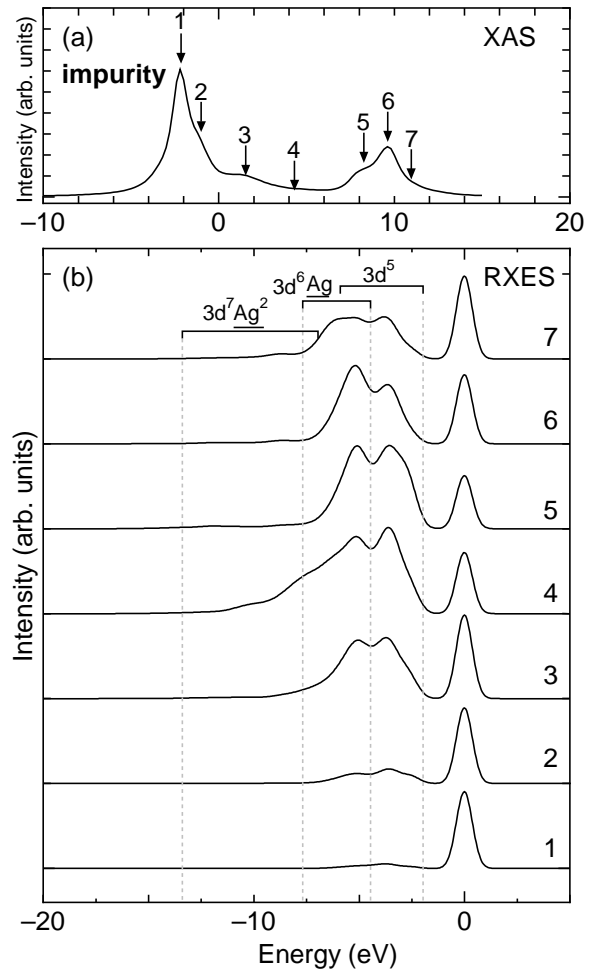}

\caption{\label{fig3} Calculated RXES spectra (b) for a Mn impurity in Ag. (a): the absorption spectrum in the Mn $L_{2,3}$ region with the indication of the excitation energies used in the RXES calculation. All RXES spectra have been rescaled to the same amplitude.}
\end{center}
\end{figure}

\section{Calculated Results}

We recall that we are considering the following three systems:
(i) a Mn impurity in Ag (labelled "impurity" in the figures below)
(ii) an adsorbed Mn monolayer on Ag (labelled "ML"),
and (iii) a thick Mn film (labelled "bct").

A schematic energy diagram of the initial (and RXES final) states for
the three considered systems is shown in Fig.~2. The configuration averaged
energies of the ionic configurations have been used which means that
multiplet, crystal field and hybridization effects are neglected in
the diagram, which would otherwise become too complex. It should be
kept in mind, however, that all these effects shift and spread the
levels. If we had represented, for example, the lowest lying multiplet
energy of each configuration instead of the configuration averaged
energies, the level ordering would be changed in several cases. In
particular, the ($3d^5$ $^6$S) term is lowest in energy for all systems
including bct Mn and ML rather than a $3d^6$($\underline{\rm Mn^+}$) term as
suggested by the configuration averaged energies. Despite these
subtleties, it is instructive to look at this simplified diagram in
order to get an idea about the CT excitation energies in the different
systems.

\begin{figure}
\begin{center}
\includegraphics[scale=.50]{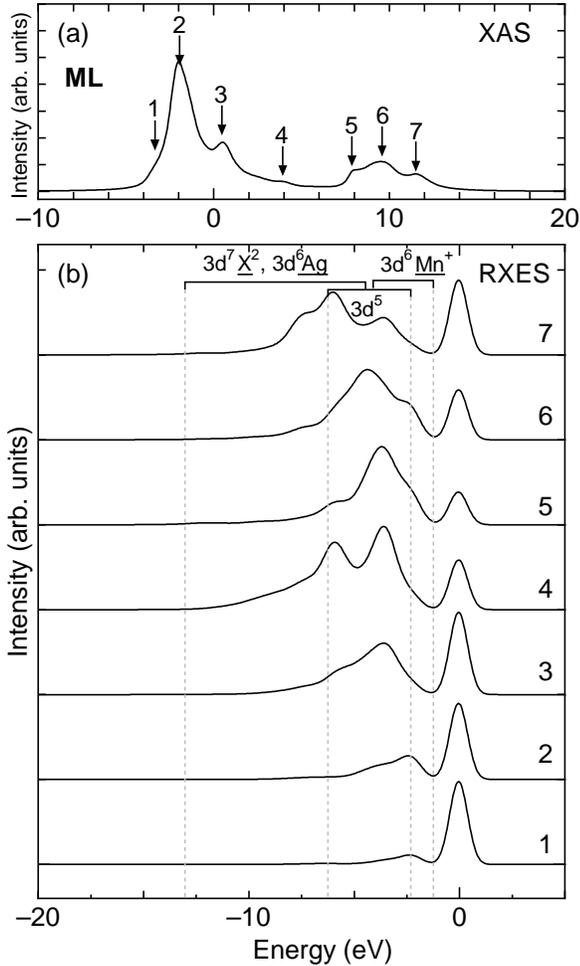}

\caption{\label{fig4} Calculated RXES spectra (b) for a Mn monolayer on Ag. (a): the absorption spectrum in the Mn $L_{2,3}$ region with the indication of the excitation energies used in the RXES calculation. $X$ stands for Mn$^+$ and Ag. All RXES spectra have been rescaled to the same amplitude.}
\end{center}
\end{figure}

Let us start with the simplest system considered here:
a Mn impurity in Ag. 
The calculated Mn $2p$ XAS is shown in Fig.~3(a).
Note that in contrast to Ref.~\cite{tag04} we have now used an
artificially reduced core hole lifetime
broadening $\Gamma_L$ of 0.3 eV. This was done merely to enhance the
resolution of spectral features for easier comparison.
The calculated Mn $2p\to3d\to2p$ RXES spectra are shown in Fig.~3(b) at various excitation energies across the $L_{2,3}$ energy region shown by arrows on the XAS spectra in Fig.~3(a). The Gaussian broadening is taken to be $\sigma =$0.5 eV.
Two main inelastic peaks can be seen around, respectively,
4 eV and 5.5 eV below the elastic peak.
At excitation energies on the $L_3$ white line (labelled 1 and 2),
the elastic peak clearly dominates.
For excitation energies increasing from $L_3$ to $L_2$ (excitation 3-5),
most of the spectral weight shifts to the inelastic peaks
before it goes partly back to the elastic one (exciatitation 6-7).
On the spectra 4-7 a weak and broad
feature at 7-13 eV can also be distinguished.

\begin{figure}
\begin{center}
\includegraphics[scale=.50]{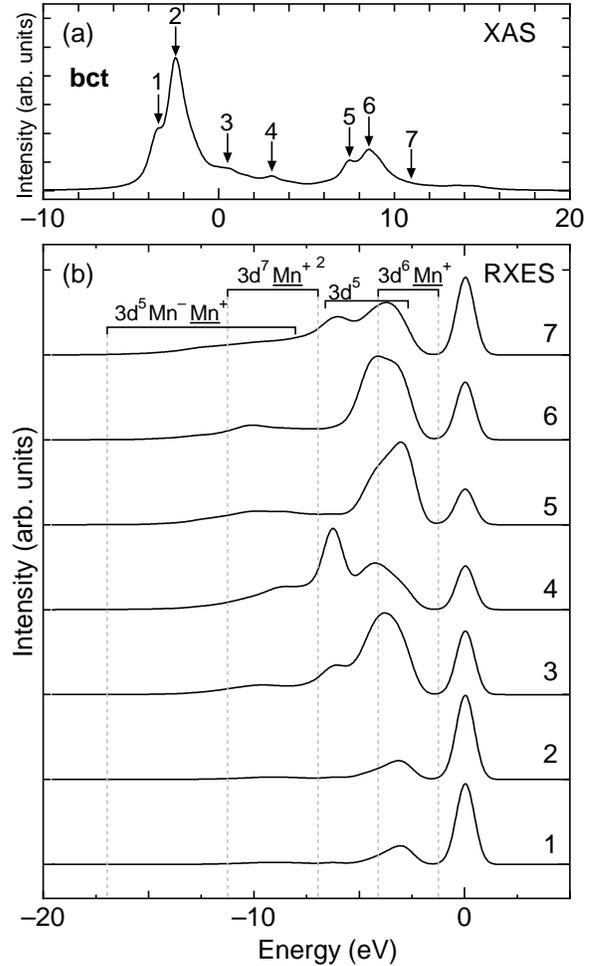}

\caption{\label{fig5} Calculated RXES spectra (b) for bulk bct Mn. (a): the absorption spectrum in the Mn $L_{2,3}$ region with the indication of the excitation energies used in the RXES calculation. All RXES spectra have been rescaled to the same amplitude.}
\end{center}
\end{figure}

The RXES features are divided into three groups named as $3d^5$, $3d^6(\underline{\rm Ag})$ and $3d^7(\underline{\rm Ag})^2$ as is depicted in the left panel in Fig.~2 and also in Fig.~3(b). The enhancement of $3d^5$ state mainly occurs well above the $L_3$ edge region (from excitations 3 to 7) with an intensity maximum just in between
the $L_3$ and $L_2$ edges (excitations 4 and 5).
This feature is assigned to the on-site $dd$ excitations in the
$3d^5$ configuration.
On the other hand, with further increasing the excitation energy, the $3d^6 (\underline{\rm Ag})$ feature is enhanced and has maximum intensity at
excitation 4-7.
Since the $3d^7(\underline{\rm Ag})^2$ configuration lies far above $3d^5$ and $3d^6(\underline{\rm Ag})$
(see Fig.~2), it has a very small weight in the ground state,
which explains the weakness of the corresponding feature in the RXES.

\begin{figure}
\begin{center}
\includegraphics[scale=.50]{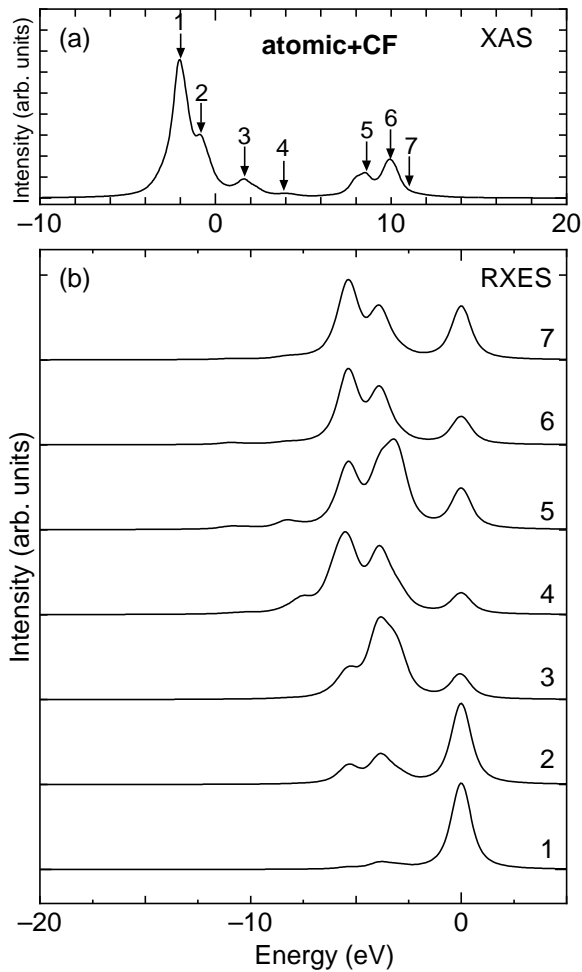}

\caption{\label{fig6} RXES spectra (b) from atomic multiplet calculation with crystal field for the 3d$^5$ configuration. (a): the absorption spectrum in the Mn $L_{2,3}$ region with the indication of the excitation energies used in the RXES calculation. All RXES spectra have been rescaled to the same amplitude.}
\end{center}
\end{figure}

\begin{figure}
\begin{center}
\includegraphics[scale=.50]{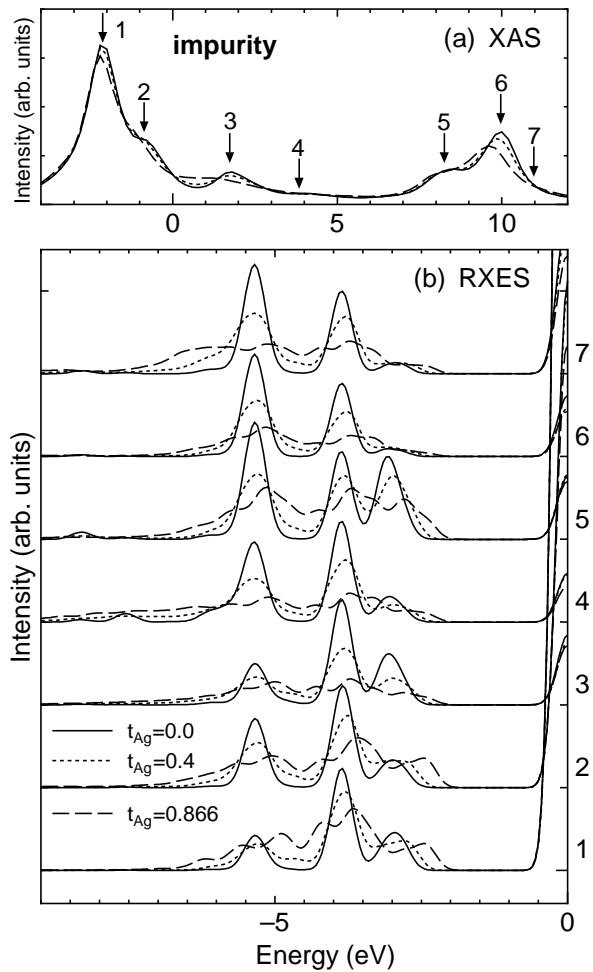}

\caption{\label{fig7} Calculated RXES spectra (b) as a function of hybridization $t_{\rm Ag}$ for a Mn impurity in Ag. (a): the absorption spectrum in the Mn $L_{2,3}$ region with the indication of the excitation energies used in the RXES calculation. RXES spectra are presented with a somewhat artificially reduced broadening in order to show some fine structure. All RXES spectra are normalized for the maximum intensity of the inelastic satellites.}
\end{center}
\end{figure}

Next we consider the monolayer (ML) system. Since CT can occur both
from the Ag and the Mn$^+$ band, its electronic structure is more
complicated than the "impurity" system (see the middle panel in Fig.~2).
The calculated RXES spectra are shown in Fig.~4.
The most simple interpretation of these spectra would be to superimpose
CT satellites originating from $3d^6$($\underline{\rm Mn^+}$) onto 
the "impurity" spectra with some weighting factor.
This approach neglects, however, configuration mixing in the final state,
which strongly modifies the spectral shape, especially the relative peak
intensities.  
For the RXES spectra at excitations 4-6, the inelastic peaks dominate
over the elastic one, a feature that is common to all three systems.
At excitation 3, there is one broad inelastic structure
centered at 3-4 eV with a weight comparable to the elastic peak.
This inelastic peak is mainly due to the anti-bonding state that is formed
by the hybridizaton between $3d^5$ and $3d^6(\underline{\rm Mn^+})$, together with its multiplet structure.
On the other hand, the spectra at excitation 4 has one more
inelastic peak around 6 eV which we assign to the $3d^6$($\underline{\rm Ag})$ configuration. The reason for this is that excitation 4 corresponds
to a XAS final state energy equal to that of the $2p^53d^7(\underline{\rm Ag})$ configuration.
For higher excitation energies (5-7), weak and broad structures appear
at 7-10 eV energy shift, in addition to the strong inelastic peak already
mentionned. 
These structures are mainly due to  the $3d^5 (\rm Mn^-)( \underline{\rm Mn^+})$ and $3d^7 (\underline{\rm Mn^+})^2$ states.
At excitation 7, the 3d$^6$($\underline{\rm Ag})$ peak around 6 eV
appears again for the same reason as in the case of excitation 4.

\begin{figure}
\begin{center}
\includegraphics[scale=.50]{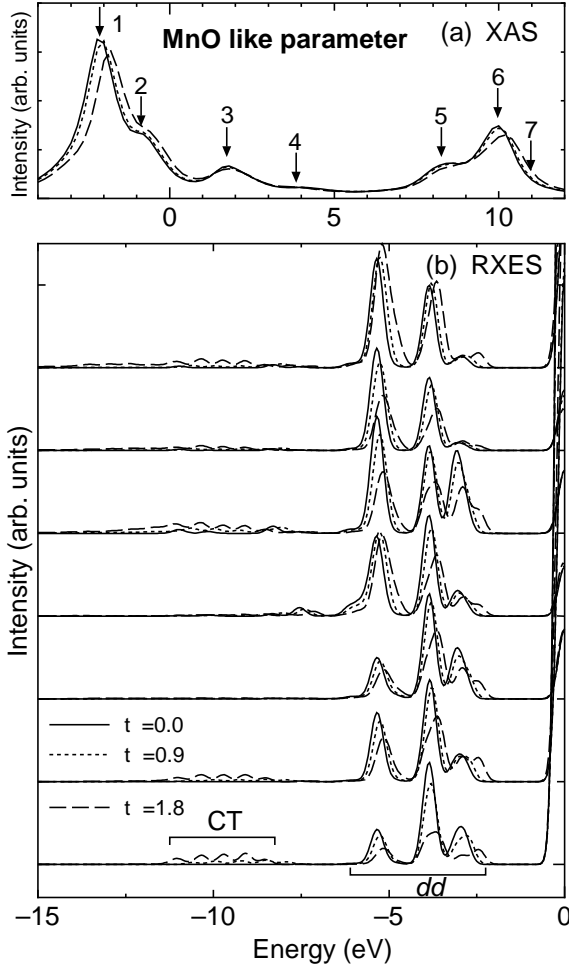}

\caption{\label{fig8}  Calculated XAS and RXES spectra for MnO-like parameter.  RXES spectra are presented with a somewhat artificially reduced broadening in order to show some fine structure. All RXES spectra are normalized for the maximum intensity of the inelastic satellites.}\end{center}
\end{figure}

Finally, we move to the bct thick films where there is no charge transfer from the Ag substrate but a considerably strong hybridization (nearly two times larger than that of the ML system) from the neighboring Mn atoms exists. 
The calculated Mn $2p\to3d\to2p$ RXES spectra for bct thick films are shown in Fig.~5. 
As depicted in the right panel in Fig.~2, the ground state is described by $3d^5$, $3d^{6}(\underline{\rm Mn^+ })$, $3d^{7}(\underline{\rm Mn^+ })^2$, $3d^4(\rm Mn^- )$, $3d^5(\rm Mn^-)(\underline{\rm Mn^+})$.
Due to the comparable energy of the $3d^5$ and $3d^{6}(\underline{\rm Mn^+ })$ states and the 
enhanced hybridization strength as compared to ML,
the ground and RXES final states have strong configuration mixing.
This results in more complicated RXES spectral shapes.
We divide the RXES features into four groups named as $3d^6(\underline{\rm Mn^+})$, $3d^5$, $3d^7(\underline{\rm Mn^+})^2$ and $3d^5(\rm Mn^-)(\underline{\rm Mn^+})$.
In contrast to a Mn impurity in Ag, the inelastic peaks with
the lowest energy shift are assigned to a CT excitation from $3d^{6}(\underline{\rm Mn^+ })$
states, rather than to a $dd$ excitation.
This is because, as we mentioned just before, the configuration-averaged energy of $3d^6 (\underline{\rm Mn^+})$ is slightly lower in energy than that of the $3d^5$ state and they are strongly hybridized with each other.

On the other hand, the $3d^5$ feature lies around 5-8 eV which is enhanced at excitation 4 and 7. Note that this $3d^5$ feature is not due to the $dd$ transitions but to anti-bonding states which are formed by hybridized $3d^6(\underline{\rm Mn^+})$ and $3d^5$ configurations. Since the Mn$^+$ bands are rather strongly hybridized with the central Mn atom, the $3d^7(\underline{\rm Mn^+})^2$ feature around 8-12 eV and broad $3d^5(\rm Mn^-)( \underline{\rm Mn^+})$ features are also seen in the RXES spectra at excitations 3, 6 and 7. 

\begin{figure}
\begin{center}
\includegraphics[scale=.60]{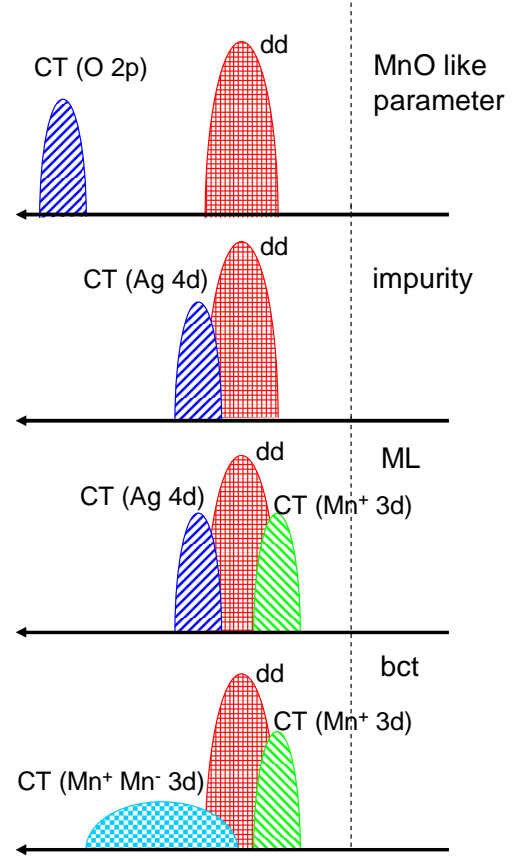}

\caption{\label{fig9} (Color online) Schematic ground states and final states of $2p\to3d\to2p$ RXES spectra in MnO, Mn impurity in Ag, ML, and bct. }
\end{center}
\end{figure}

\section{Discussion}

As shown above, the spectral shape of the inelastic satellites changes considerably among the three considered systems.  
For a Mn impurity in Ag, the two main inelastic peaks are
essentially due to $dd$ excitations, but the one at higher energy shift (5-6eV)
is modified by a CT excitation from the Ag band.
For the ML system, on the other hand, the lowest satellite is a mixture of $dd$ excitations and CT satellite from the Mn$^+$ band,
while the part with higher energy shift ($\simeq $6 eV) is affected by CT
satellites from the Ag band.
For the bct system, finally, the CT satellite from the Mn$^+$ band has
the lowest excitation energy, whereas the $dd$ structure is hard to see
in this system. 
Since the spectral shapes are considerably complicated, explaining their dependence is not a trivial task from a general point of view. 
In this section, we show the hybridization strength dependence of RXES spectra to clearly distinguish between $dd$ structure and CT satellite and to understand how CT excitations modify the $dd$ structure.

\begin{figure}
\begin{center}
\includegraphics[scale=.50]{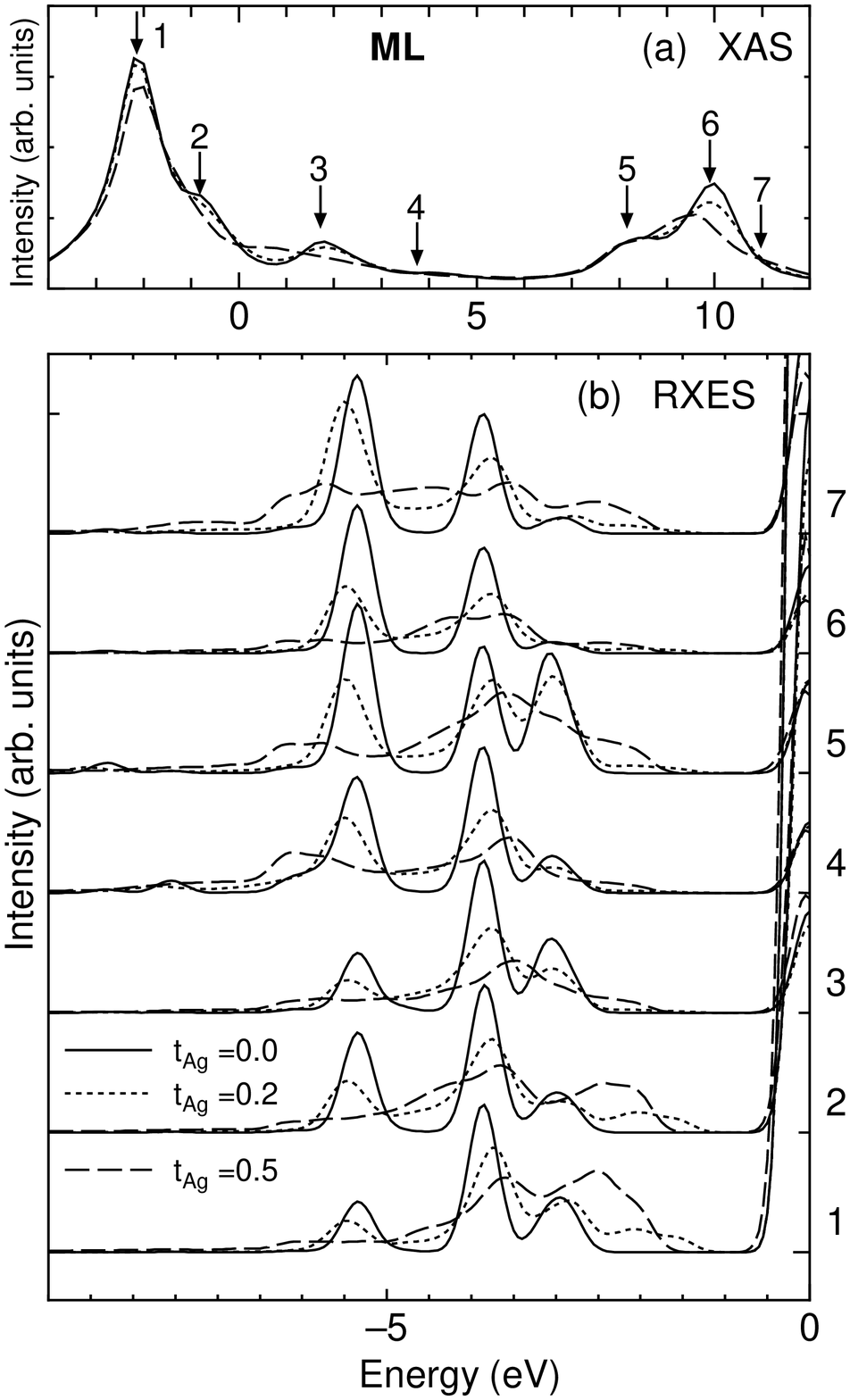}

\caption{\label{fig10} Calculated RXES spectra as a function of hybridization $t_{\rm Ag}$ with keeping the ration $t_{\rm Ag} = 5/8 t_{\rm Mn}$(b) in a wide energy window of Mn/Ag. (a): the absorption spectrum in the Mn $L_{2,3}$ region with the indication of the excitation energies used in the RXES calculation. RXES spectra are presented with artificially reduced broadening. All RXES spectra are normalized for the maximum intensity of the inelastic satellites.}
\end{center}
\end{figure}~

For this purpose, we start by considering the atomic limit to view the original $dd$ structure of $3d^5$. In Fig.~6, we have performed atomic multiplet calculations including the crystal field with O$_h$ symmetry (10Dq=0.5 eV). The $dd$ excitation has basically three structures at -3 eV, -4 eV, and -5.5 eV, while their relative intensities depend on the excitation energy.  Qualitatively, the atomic RXES spectra ressemble those of Mn impurity in Ag and ML, but there are some discrepancies. In Mn impurity in Ag, the intensity of the lower part of $dd$ structure (around 5-6 eV) is decreased and the peak become broad in contrast to the atomic calculation. In ML, new structures can be seen around 2.5 eV in addition to the lower part changes as well as the case of Mn impurity in Ag.

\begin{figure}
\begin{center}
\includegraphics[scale=.50]{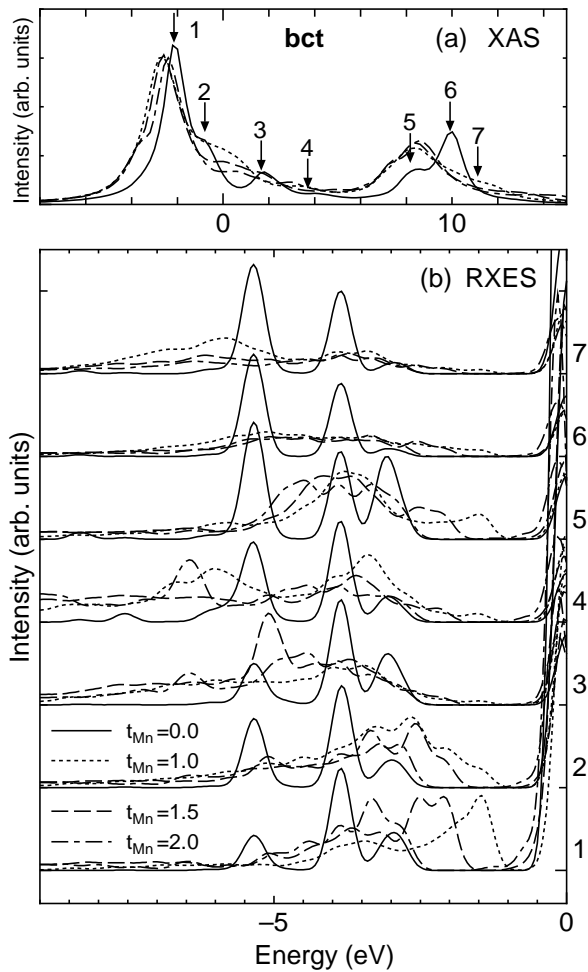}

\caption{\label{fig11} Calculated RXES spectra as a function of hybridization $t_{\rm Mn}$ (b) in a wide energy window of bct thick films. (a): the absorption spectrum in the Mn $L_{2,3}$ region with the indication of the excitation energies used in the RXES calculation. RXES spectra are presented with artificially reduced broadening. All RXES spectra are normalized for the maximum intensity of the inelastic satellites.}
\end{center}
\end{figure}

Next we study the dependence of RXES spectra as a function of hybridization. 
Figure~7 summarizes a qualitative study of the effect of increasing the hybridization $t_{Ag}$ for impurity system with respect to fixed parameters ($U_{dd}$, $U_{dc}$, $\varepsilon_{\rm Ag}$, $\Delta$ and 10Dq). 
As can be seen from Fig.~7, a CT satellite originating from $3d^6(\underline{\rm Ag})$ 
appears when switching on the hybridization to the Ag band; its
energy position almost coincides with the left-most of the three
$dd$-peaks (at 5-6 eV). This is due to the small charge transfer energy $\Delta_{\rm Ag}\equiv \Delta-\varepsilon_{\rm Ag}\approx 2.0$ eV with the notation of Ref.[2]. 

In a typical TM compound  such as
MnO, CT and $dd$ excitations are clearly separated in the RXES spectra as shown in Fig.~8\cite{but96}.
The $dd$ structure is then hardly affected by the CT satellite
even with some moderate hybridization strength. This is also indicated in
the upper part of Fig.~9.
In the present system, Mn impurity in Ag, the CT and $dd$ excitation
energy regions partly overlap (at 5-6 eV) as shown in the middle part
of Fig.~9. The two types of excitations can therefore easily mix.
This leads to a deformation of the $dd$ structure at 5-6 eV which increases
with hybridization strength.
On the other hand, the hybridization dependence of the upper part of $dd$ structure (around 3-4 eV) is similar to the crystal field dependence ($i.e.$ increasing the crystal field results in the change of the relative peak intensity and the peak shift only) and also to the hybridization dependence for MnO like parameters as shown in Fig.~8.

Let us next look at the hybridization dependence in the
ML system. We show in Fig.~10 the RXES of ML for the hybridization $t_{\rm Ag}$ varied between 0 eV and 0.5 eV with keeping the ratio $t_{\rm Ag}=5/8t_{\rm Mn}$ . 
Figure~10 clearly shows that the inelastic structures with lowest energy
shift (2-3 eV) are due to the hybridization with the Mn$^+$ bands.
Since the averaged configuration energy of $3d^{6}$($\underline{\rm Mn}$$^+$) is very close to $3d^5$ configuration ($\Delta_{\rm Mn}\approx -0.5$ eV), the corresponding CT satellite is lower than the $dd$ excitation energy as is shown in the middle panel in Fig.~9.
However, since the actual hybridization strength is not so big
in the Mn/Ag system ($t_{\rm Ag} =0.5$ eV and $t_{\rm Mn} = 0.8$ eV) the $dd$ structure can still be recognized in the spectrum.

Finally, let us consider the $t_{\rm Mn}$ dependence of RXES spectra for bct thick film. As shown in Fig.~11,  the RXES spectra strongly depends on $t_{\rm Mn}$. This is partly due to the fact that the hybridization strength is much
bigger than in ML. In this case the original $dd$ peaks are
drastically modified and the spectra have a much richer structure.

We have shown that CT screening from the Ag and Mn bands strongly
modifies the $2p\to3d\to2p$ RXES line shape. In contrast to typical TM
compounds, there is a strong mixing between $dd$ excitations
and CT satellites, which is mainly due to the much smaller CT energies
involved.

\section{Concluding Remarks}

In summary we have presented calculations of Mn $2p$ XAS and $2p\to3d\to2p$ RXES
spectra for three kinds of Mn/Ag thick film structure. We have
used an impurity model including full multiplet interaction
and coupling to the Mn $3d$ and Ag $4d$ bands.
This allowed us to investigate the interplay between on-site $dd$
excitations and charge transfer screening from neighboring Mn and
Ag atoms. 
The calculated RXES line shapes are very sensitive to the
hybridization strength of Mn $3d$ and Ag $4d$ orbitals.
Since our results show that in systems with small charge transfer energy 
the $dd$ excitation structure is considerably modified by charge
transfer screening, it would be interesting to verify it experimentally.

\section*{Acknowledgment}

Financial support by the Centre National de la Recherche Scientifique through the a PICS program between France and Japan is gratefully acknowledged.

\end{document}